\documentclass[a4paper,aps,pre,superscriptaddress,floatfix,nofootinbib,longbibliography,notitlepage,onecolumn]{revtex4-1}
\usepackage{bbm}
\usepackage{bbold}
\usepackage{bbm}
\usepackage{longtable}
\usepackage[pdftex]{graphicx}
\usepackage{latexsym,amsmath,verbatim,amssymb,txfonts}
\usepackage{lipsum}
\usepackage{threeparttable}
\usepackage{tabularx}
\usepackage{booktabs}
\usepackage{color}
\usepackage{rotating}
\usepackage{verbatim}
\usepackage{multirow}
\usepackage[english]{babel}
\usepackage{comment}
\usepackage{bm}
\usepackage{amsmath}
\usepackage{amsfonts}
\usepackage{amssymb}
\usepackage{color}
\usepackage{braket}
\usepackage{blkarray, bigstrut}

\begin{document}
\title{Theory of phase reduction from hypergraphs to simplicial complexes: a general route to higher-order Kuramoto models}

\author{Iv\'an Le\'on}
\email{leonmerinoi@unican.es}
\affiliation{Department of Applied Mathematics and Computer Science, Universidad de Cantabria, Santander, Spain}

\author{Riccardo Muolo}
\affiliation{Department of Systems and Control Engineering, Institute of Science Tokyo (former Tokyo Tech), Tokyo 152-8552, Japan}

\author{Shigefumi Hata} 
\affiliation{Graduate School of Science and Engineering, Kagoshima University, Kagoshima 890-0065, Japan}

\author{Hiroya Nakao}
\affiliation{Department of Systems and Control Engineering, Institute of Science Tokyo (former Tokyo Tech), Tokyo 152-8552, Japan}
\affiliation{International Research Frontiers Initiative, Institute of Science Tokyo (former Tokyo Tech), Kanagawa 226-8501, Japan}

\date{\today}

\begin{abstract}

{Kuramoto-type models are paradigmatic models
in the study of coupled oscillators, since they provide a simple explanation of the collective synchronization transition. Their universality comes from the fact that they are derived through phase reduction. Recent evidence highlighting the importance of higher-order (many-body) interactions in the description of real-world systems has led to extensions of Kuramoto-type models to include such frameworks.
However, most of these extensions were obtained by adding higher-order terms instead of performing systematic phase reduction, leaving the questions of which higher-order couplings naturally emerge in the phase model. In this paper, we fill this gap by presenting a theory of phase reduction for coupled oscillators on hypergraphs, i.e., the most general case of higher-order interactions. We show that, although higher-order connection topology is preserved in the phase reduced model, the interaction topology generally changes,
due to the fact that the hypergraph generally turns into a simplicial complex in the reduced Kuramoto-type model. Furthermore, we show that, when the oscillators have certain symmetries, even couplings are irrelevant for the dynamics at the first order. The power and ductility of the phase reduction approach are illustrated by applying it to a population of Stuart-Landau oscillators with an all-to-all configuration and with a ring-like hypergraph topology; in both cases, 
the analysis of the phase model can provide deeper insights and analytical results.}

\end{abstract}

\maketitle

\section{Introduction}

The understanding, prediction, and control of the dynamics of populations of nonlinear oscillators is a fundamental topic in several fields, from biology~\cite{Win80,GM88} and engineering~\cite{,frasca2018synchronization}, to physics~\cite{PRK01} and chemistry~\cite{kuramoto1975}, to name a few.
{The intrinsic} complexity of those systems {hinders} the understanding of
their dynamics, so simplification and reduction methods are in order. { When the interactions among the units are weak, it is convenient to apply the phase reduction~\cite{kuramoto1975,nakao16}, which is a perturbative technique that captures the dynamics of each oscillator solely in terms of one variable: the phase $\theta$. The low dimensionality and simplicity of the phase models allows a better understanding of the dynamics.} In fact, the Kuramoto model 
is derived through phase reduction{\footnote{{Originally by adiabatic elimination in~\cite{kuramoto1975}, and then systematically by phase reduction in~\cite{Kur84}.}}}, providing a simple and general mechanism for the emergence of synchronization~\cite{kuramoto1975,Kur84}. 
In recent years, extensions of this method have been considered such as analyzing 
higher-order corrections in the perturbative approach~\cite{leon19,leon22a,Mau23,bick2024higher} or the dynamics of the amplitude variables under non-weak coupling~\cite{kurebayashi2013phase,Wilson_2016,shirasaka2017phase,leon20}.

In complex systems, the dynamics are shaped by the interactions among the elementary units and such interactions have 
mostly been assumed to be pairwise (two-body). However, in recent years, increasing evidence emerged towards the fundamental role that higher-order (many-body) interactions play in describing real-world systems~\cite{battiston2020networks,bianconi2021higher,kuehn2021universal,bick2023higher,boccaletti2023structure,Herinrich11,Matthew19}. In fact, it has been shown that higher-order interactions enrich the dynamics of the systems, with applications regarding synchronization~\cite{leon22a,skardal2020higher,millan2020explosive,lucas2020multiorder,gambuzza2021stability}, random walks~\cite{schaub2018random,carletti2020random}, pattern formation~\cite{carletti2020dynamical,muolo2024turing}, opinion dynamics~\cite{iacopini2019simplicial,neuhauser2020multibody}, and pinning control~\cite{de2022pinning,xia2024pinning}, to name a few. In all the above works, the authors showed that certain dynamics are only made possible by complex interactions described by a higher-order topology, namely, hypergraphs and simplicial complexes, which are extensions of pairwise networks. This is particularly interesting, as the complexity of the system lies not in the model, but rather in the interactions between the elementary units, and even very simple settings yield rich dynamics when higher-order interactions are present{, as was shown by applying phase reduction to a system of globally (all-to-all) coupled Stuart-Landau oscillators with $2$- and $3$-body interactions~\cite{leon24higher}.}

{ However, a general theory of phase reduction for arbitrary higher-order coupling is yet to be developed, and this work aims to fill the gap. In fact, the goal of this paper is to present 
a general phase reduction method for nonlinear oscillators interacting 
through a hypergraph, 
the most general representation of higher-order interactions. The phase reduction method provides three advantages: a) it yields lower-dimensional models with higher symmetry that are easier to analyze; b) it 
reveals which interactions terms naturally emerge in the phase 
(Kuramoto-type) models; c) it 
justifies the study of those phase models 
as models of real-world systems, as they capture the dynamics of real populations of oscillators (when weakly coupled). We remark that phase reduction in presence of higher-order interactions has been previously performed, but always for specific examples of oscillators with rotational symmetry and either all-to-all coupling~\cite{ashwin2016hopf,leon24higher}
or a 
specific uniform topology~\cite{muolo2024pinning} has been assumed (see also the appendix of \cite{KomPik11} where the functional form of a phase reduced model for two non-resonant 
globally coupled oscillators was derived). In this work, we present 
a general theory, applicable to any system of oscillators coupled via any hypergraph. Some general conclusions will be derived from this method, such as that the adjacency tensor (connection topology) is preserved after the reduction, though the interaction topology, in general, changes due to the emergence of simplicial complexes, 
and that, under certain symmetries of the oscillator, even couplings (as quadratic or quartic) do not contribute to the dynamics.

The applicability of the theory and advantages of the phase reduced models are pedagogically exemplified for the Stuart-Landau (SL) oscillator, where the rotational symmetry allows to obtain further analytical results. In this example, the ductility of the theory is obvious, since it allows a detailed analysis of the dynamics of the populations of SL oscillators through the study of their reduced phase model, which is difficult
through 
other methods, such as the Master Stability Function (MSF)~\cite{fujisaka1983stability,pecora1998master}
; see~\ref{app:a} for details.}

The paper is organized as follows. Sec. II is devoted to a classical introduction of phase reduction, starting from the definition of phase. In Sec. III we present the theory to take into account higher-order interactions. We exemplify the usefulness of the theory by applying it to SL oscillators in Sec. IV, where multiple types of couplings are considered and 
analyzed. Finally, in Sec. V, we include the conclusions, discussion, and implications of this work.

\section{Classical phase reduction theory}\label{sec:phasered}

In this section, we present { an introduction to} the classical theory of phase reduction~\cite{Kur84,KuraNakao19}. Phase reduction is a dimensionality reduction method that allows to describe the dynamics of each oscillator in terms of a single phase variable. This is done through a perturbative expansion, assuming a small coupling strength among the units. { This powerful technique provides a solid theoretical foundation for the use of phase models in studying the dynamics of coupled oscillators. Under the assumption of sufficiently weak coupling, any ensemble of interacting oscillatory units can be approximated by a phase model. These reduced models, due to their simplicity, facilitate deeper insights into collective behavior, as illustrated by the Kuramoto model, providing a simple mechanism for collective synchronization.}

Phase reduction is applicable whenever we have a system possessing an attractive limit cycle subject to weak perturbations or interactions:
    \begin{equation}\label{eq.genosc}
        \dot{\bm{X}}=\bm{F}(\bm{X})+\epsilon \bm{g}(\bm{X},t),
    \end{equation}
where $\bm{X}(t)$ 
is the 
state variable
of the oscillator{\footnote{{For simplicity, we assume the oscillator to be finite dimensional. Note, however, that phase reduction can be extended to infinite-dimensional systems, as done in \cite{KawaNaka08,Nakao14}.}}}, $\bm{F}$ is the velocity vector field, responsible for an exponentially stable limit cycle solution when $\epsilon=0$, $\epsilon$ 
is a small parameter characterizing the strength of the perturbation, and $\bm{g}$ is the external perturbation. Note that the time dependence of $\bm{g}$ can be due to the effect of a forcing or to the interaction with other oscillators. 

\subsection{Isolated oscillator, $\epsilon=0$}

\begin{figure}[h!]
\includegraphics[width=\linewidth]{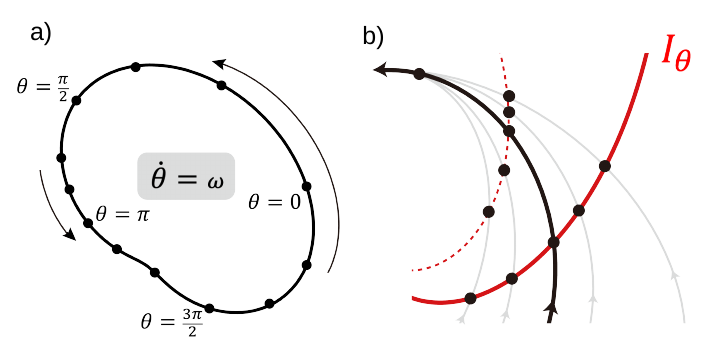}
	\caption
	{Schematic representation of the definition of phase on the limit cycle (a) and the asymptotic phase and isochrons (b).}
	\label{fig:Fig1}
\end{figure}

In the absence of 
perturbations, i.e., $\epsilon=0$, the system evolves into a limit cycle solution with period $T$, denoted by $\bm{X}_c(t)=\bm{X}_c(t+T)$. On the limit cycle, we can define the phase of the oscillator as a variable whose rate of increase is always equal to the natural frequency 
$\omega = 2\pi / T$ of the oscillator:
\begin{equation}
    \dot{\theta}=\omega,
\end{equation}
which means that every period the phase advances $2\pi$. Such concept is schematically depicted in Fig. \ref{fig:Fig1}(a).

We can then extend this definition using the concept of asymptotic phase. We say that a point in the basin of attraction of the limit cycle $\bm{X}_1(0)$ has the same phase as $\bm{X}_2(0)$ lying on the limit cycle, if $\bm{X}_1(t)$ converges to 
$\bm{X}_2(t)$
as $t \to \infty$.  This definition foliates the basin of attraction in isochrons $I_\theta$, level sets of the asymptotic phase, with $\theta \in [0, 2\pi)$;
all the points belonging to the same isochron $I_\theta$ have the same phase $\theta$, as represented in Fig. \ref{fig:Fig1} (b).

The definition of isochrons ensures that every point in the basin of attraction has assigned the phase of its isochron, $\theta(\bm{X})=I_\theta(\bm{X})$. 
From the definition of the phase, it is straightforward to prove that
\begin{equation}
    \dot{\theta}=\nabla_{\bm{X}}\theta\cdot\frac{d\bm{X}}{dt}=\nabla_{\bm{X}}\theta \cdot \bm{F}(\bm{X})=\omega,
\end{equation} which yields the trivial evolution of the phase.

\subsection{Perturbed oscillator, $\epsilon\not=0$}

We are now interested in the effect of the 
perturbation on the phase. Let us consider the weak perturbation
$\epsilon \bm{g}(\bm{X},t)$ applied to our oscillator. 
The evolution equation 
for the phase is derived 
by applying the chain rule to the definition of phase and Eq.~\eqref{eq.genosc}.
\begin{equation}
    \dot{\theta}=\nabla_{\bm{X}} \theta(\bm{X}) \cdot \frac{d\bm{X}}{dt}
    =\nabla_{\bm{X}}\theta(\bm{X}) \cdot \bm{F}(\bm{X})+\epsilon \nabla_{\bm{X}} \theta(\bm{X}) \cdot \bm{g}(\bm{X},t).
\end{equation}
Employing the relation $\nabla_{\bm{X}}\theta(\bm{X}) \cdot \bm{F}(\bm{X})=\omega$ yields:
\begin{equation}\label{eq.phaseopen}
    \dot{\theta}=\omega+\epsilon \nabla_{\bm{X}} 
    \theta(\bm{X}) \cdot \bm{g}(\bm{X},t).
\end{equation}

Equation~\eqref{eq.phaseopen} is exact, no approximation has been done yet; however, it is not a closed expression since there is still a dependence on $\bm{X}$. Thus, no dimensionality reduction has been 
achieved yet. The crucial step of phase reduction theory is the assumption of $\epsilon/\lambda\ll1$,
where $\lambda$ is the second Floquet exponent
characterizing the relaxation rate of the oscillator state to the limit cycle\footnote{The first Floquet exponent can be taken as $0$ or $i \omega$, which corresponds to the neutral direction along the limit cycle.}. In that case, the timescale in which the oscillator decays to the limit cycle is much faster than the change in the phase, implying that the oscillator remains always close to the limit cycle. Thus, its 
state is $\bm{X}=\tilde{\bm{X}}_c(\theta)+O(\epsilon/\lambda)$, where $\tilde{\bm X}_c(\theta) = {\bm X}_c(\theta / \omega)$ is a state on the limit cycle with phase $\theta$. {For the sake of notation and without loss of generality, we fix $\lambda=1$ from now on (this is possible 
by time rescaling)}. We can now plug the above expression 
into Eq. \eqref{eq.phaseopen}, obtaining:
\begin{equation}\label{eq.pheq}
    \dot{\theta}=\omega+\epsilon \bm{Z}(\theta) \cdot \bm{g}(
    \tilde{\bm X}_c(\theta),
    t)+O(\epsilon^2),
\end{equation}
where we have defined 
$\bm{Z}(\theta)=\nabla_{\bm{X}} \theta |_{{\bm X} = \bm{\tilde{X}}_c(\theta)}$,
called phase sensitivity function\footnote{Such expression is 
also called \textit{infinitesimal phase resetting curve} or \textit{phase response function} (but note that $\bm{Z}$ represents the linear response coefficient, not the phase response itself).}, which measures the change rate of the phase by an infinitesimal perturbation.

It is relevant to notice that phase reduction theory is, in general, performed by neglecting terms of order $\epsilon^2$. These terms, however, might have importance qualitatively and quantitatively. For a derivation of these terms and its effects, see~\cite{KuraNakao19,leon19,leon22a,Mau23,mau2024phase,bick2024higher,Herinrich11,Matthew19}. 

If the difference between the natural frequency of the oscillator and the frequency of the interaction term is small, we can further simplify the system employing the averaging method{~\cite{PRK01}, as we show below}.

\subsection{Pairwise interactions}

We are often interested 
in the dynamics of a population of coupled oscillators. Assuming that the interactions between oscillators are pairwise, i.e., two-body, their dynamics can be written as:
\begin{equation}
    \dot{\bm{X}_j}=\bm{F}_j(\bm{X}_j)+\epsilon \sum_{k=1}^N A_{jk} \bm{g}_{jk}(\bm{X}_j,\bm{X}_k),
\end{equation}
where ${\bm F}_j$ is the vector field of oscillator $j$, $A_{jk}$ is the adjacency matrix of the interaction network \cite{Latorabook,newmanbook2}, and $\bm{g}_{jk}(\bm{X}_j,\bm{X}_k)$ is the interaction from oscillator $k$ to oscillator $j$.
We assume that ${\bm F}_j = {\bm F} + O(\varepsilon)$, i.e., all the oscillators have nearly identical properties, and use the limit cycle ${\bm X}_c$ of ${\bm F}$ as the reference orbit for phase reduction.

Proceeding analogously to the previous section, phase reduction yields:
\begin{equation}\label{eq:phase_model_pairwise}
    \dot{\theta}_j=\omega_j+\epsilon \bm{Z}(\theta_j) \cdot \sum_{k=1}^{N}A_{jk}\bm{p}_{jk}(\theta_j,\theta_k)+O(\epsilon^2),
\end{equation}
where $\bm{p}_{jk}(\theta_j,\theta_k)=\bm{g}_{jk}(\bm{\tilde{X}}_c(\theta_j),\bm{\tilde{X}}_c(\theta_k))$.

If the natural frequencies of all oscillators are close to $\omega$, i.e., $\Delta\omega_j=\omega_j-\omega=O(\epsilon)$,
we can perform the averaging procedure as follows. We introduce a transformation to the slow variable, $\phi_{j} = \theta_{j} - \omega t$, and rewrite Eq.~\eqref{eq:phase_model_pairwise} as
 \begin{equation}
    \dot{\phi}_j = \Delta\omega_j + \epsilon \sum_{k=1}^{N} A_{jk} \bm{Z}(\phi_j + \omega t) \cdot \bm{p}_{jk}(\phi_j+\omega t, \phi_k+\omega t)
     + O(\epsilon^2),
\end{equation}
where the terms on the right-hand side are of $O(\epsilon)$.
%
Thus, $\phi_{j}(t)$ are slow variables, and they remain almost constant over one cycle. This means that we can average the time-dependent terms on the right-hand side over one period of the cycle, $T=2\pi / \omega$, neglecting additional $O(\epsilon^2)$ terms~\cite{Kur84,Guckenheimer,Sanders83}, yielding
\begin{align}
    \Gamma_{jk}(\phi_k-\phi_j)
    &=\frac{1}{T}\int_0^{T}\bm{Z}(\phi_j + \omega t) \cdot \bm{p}_{jk}(\phi_j + \omega t, \phi_k + \omega t) dt \cr
    &=\frac{1}{T}\int_0^{T}\bm{Z}(\omega t) \cdot \bm{p}_{jk}( \omega t,\phi_k - \phi_j + \omega t) dt \cr
    &=\frac{1}{2\pi}\int_0^{2\pi}\bm{Z}(\varphi) \cdot \bm{p}_{jk}(\varphi,\phi_k - \phi_j + \varphi) d\varphi,
\end{align}
where we used $2\pi$-periodicity of $\bm{Z}$ and $\bm{p}_{jk}$. Thus, we approximately obtain
 \begin{equation}
    \dot{\phi}_j = \Delta \omega_j + \epsilon \sum_{k=1}^{N} A_{jk} \Gamma_{jk}(\phi_k - \phi_j),
\end{equation}
and, in the original phase variables,
 \begin{equation}
    \dot{\theta}_j = \omega_j + \epsilon  \sum_{k=1}^{N}A_{jk}\Gamma_{jk}(\theta_k-\theta_j).
\end{equation}
It is important to note that $O(\epsilon^2)$ terms have been neglected in both the phase reduction and the averaging procedure. { We remark that the averaging procedure can also be applied if the frequencies fulfill some resonant condition, such as $\omega_j\simeq 2\omega_k$, by changing the definition of the slow variables $\phi_j$ \cite{PRK01}. }

Through the phase reduction and the averaging process, all the details of the model are captured by the function $\Gamma$, which solely depends on the phase difference. If we assume that $\Gamma$ contains only the first harmonic, we obtain the celebrated Kuramoto model~\cite{Kur84}.

\section{Phase reduction with higher-order (many-body) interactions}

In the past, the coupling in ensembles of oscillators has 
mainly been considered to 
take the form of pairwise interactions. However, recent findings have shown that higher-order interaction may play a critical role in the dynamics, with particular interest for phase models~\cite{tanaka2011multistable,ashwin2016hopf,bick2016chaos,skardal2020higher,lucas2020multiorder,wang2021collective,kovalenko2021contrarians,adhikari2023synchronization,skardal2023multistability,leon24higher,costa2024bifurcations,huh2024critical,wang2024coexistence,zhang2024deeper}. { When higher-order interaction are present, the system is no longer described by a network of nodes and links, but by a hypergraph, in which interacting nodes are connected via hyperedges. A $1$-hyperedge encodes a $2$-body interaction, i.e., a link\footnote{{Note that, in some works on hypergraphs, the notation is different, namely, $m$-hyperedges encode $m$-body interactions. However, our notation is consistent with the literature on simplicial complexes, where hyperedges are called \textit{simplices}, 
a $1$-simplex is a link, a $2$-simplex is a triangle, an $m$-simplex encodes a $(m+1)$-body interaction, etc., and this is motivated by the connections with topology and exterior calculus \cite{bianconi2021higher}.}}. For $m>1$, $m$-hyperedges encode $(m+1)$-body interactions and are, hence, a generalization of the links. A particular type of hypergraph is a simplicial complex, in which if some nodes share a $m$-body interaction (i.e., a $(m-1)$-hyperedge),  all sub-interactions are present, i.e., they are also connected by all $(n-1)$-hyperedges with $n<m$. Stated differently, a hypergraph allows higher-order interactions of any order without constraint, while a simplicial complex is more restriced since if $m$ nodes share a $(m-1)$-hyperedge, they also have to be connected through all possible $n$-hyperdges with $n<m$.

In this section, we show how and which higher-order interaction terms naturally emerge when phase reduction is performed.} For the sake of simplicity, we hereby consider the pure $3$-body case, i.e., only $3$-body interactions are present, and apply the phase reduction as in the above Sec.~\ref{sec:phasered}. The extension to other (or mixed) higher-order interactions follows straightforwardly from the theory below, and all conclusions obtained are equivalent. 

\subsection{Phase reduction}

Consider a population of oscillators coupled through $3$-body interactions:
\begin{equation}
       \dot{\bm{X}}_j=\bm{F}_j(\bm{X}_j)+\epsilon \sum_{k,l=1}^{N} A_{jkl} \bm{g}_{jkl}(\bm{X}_j,\bm{X}_k,\bm{X}_l),
\end{equation}
where $A_{jkl}$ is the adjacency tensor encoding the $3$-body interaction {weights}, i.e., the hypergraph topology\footnote{As for the case of pairwise interactions, here we assume for simplicity the higher-order interactions to be symmetric, but in general also hypergraphs can be asymmetric \cite{gallo1993directed,gallo2022synchronization}.}. The only difference from the pairwise case is that now the interaction function $\bm{g}_{jkl}$ depends on three variables. We can then apply the previous theory, obtaining the following phase reduced equation:
\begin{equation}
    \dot{\theta}_j=\omega_j+\epsilon \bm{Z}(\theta_j) \cdot \sum_{k,l=1}^{N} A_{jkl} \bm{p}_{jkl}(\theta_j,\theta_k,\theta_l) + O(\epsilon^2),
\end{equation}
where all functions have been evaluated on the limit cycle, namely, 
\begin{equation}\label{eq.gtop}
    \bm{p}_{jkl}(\theta_j,\theta_k,\theta_l)=\bm{g}_{jkl}(\bm{\tilde{X}}_c(\theta_j),\bm{\tilde{X}}_c(\theta_k),\bm{\tilde{X}}_c(\theta_l)).
\end{equation}

When all the oscillators have similar frequencies, i.e., $\omega_j=\omega+O(\epsilon)$, we can perform the averaging procedure as in the pairwise interactions case, obtaining:
\begin{equation} \label{eq.phaseredhighorder}
    \dot{\theta}_j=\omega_j+\epsilon  
    \sum_{k,l=1}^{N}
    A_{jkl}
    \Gamma_{jkl}
    (\Delta\theta_{kj},\Delta\theta_{lj})
     + O(\epsilon^2),
\end{equation}
where $\Delta\theta_{mn}=\theta_m-\theta_n$ and
the phase coupling function $\Gamma_{jkl}$ is calculated as
\begin{eqnarray}\label{eq.gammagen}
    \Gamma_{jkl}(\Delta\theta_{kj},\Delta\theta_{lj})
    &=&\frac{1}{T}\int_0^{T}\bm{Z}(\phi_j+\omega t) \cdot \bm{p}_{jkl}(\phi_j + \omega t, \phi_k + \omega t, 
    \phi_l + \omega t)dt \cr
    &=&\frac{1}{T}\int_0^{T}\bm{Z}(\omega t) \cdot \bm{p}_{jkl}(\omega t, \phi_k - \phi_j + \omega t, \phi_l - \phi_j + \omega t)dt \cr
    &=&\frac{1}{2\pi}\int_0^{2\pi}\bm{Z}(\varphi) \cdot \bm{p}_{jkl} (\varphi,\Delta \theta_{kj} + \varphi, \Delta \theta_{lj} + \varphi)d\varphi ,
\end{eqnarray}
where  
$\phi_{i} = \theta_{i} - \omega t$ 
as 
before { and 
we made the variable change $\varphi=\omega t$}
in the last line.
This $\Gamma_{jkl}$ {function} is $2\pi$-periodic with respect to each of its arguments. {The averaging procedure can be implemented for other resonance conditions $m \omega_j- n\omega_l=O(\epsilon)$ \cite{PRK01}, or even in non-resonant cases \cite{KomPik11}.
Although the different frequency relations yield different dynamics~\cite{Dai2025},
in many real-world systems, the units are similar and thus have close frequencies, making our consideration $\omega_i=\omega+O(\epsilon)$ rather general.}\\

{ From Eq.~\eqref{eq.phaseredhighorder}, some conclusions can be drawn:}
\begin{itemize}
    \item The adjacency tensor, i.e., the { connection topology}, of the phase oscillators is identical to that of the non-reduced system at the first order phase reduction performed here. {Notice that this does not imply that the interaction topology is preserved, since the interaction is mediated by $\Gamma$. If the function $\Gamma$
    vanishes, the corresponding hyperedge disappears. Similarly, if it only depends on one of the phase differences (e.g., $\Delta\theta_{kj}$), the three-body interaction would become a pairwise interaction, see Eq. \eqref{eq.phaseredhighorder}. Notice that, differently to the second order phase reduction~\cite{leon19,Mau23,bick2024higher}, interactions involving more than three bodies or among units that are not initially connected cannot emerge.  In general, if the connection topology contains $n$-body interactions, then,
    all $m$-body interactions with $m\leq n$ are part of the interaction topology of the phase reduced model. 
    In that sense, the original hypergraph becomes a simplicial complex. In Section~\ref{sec:degrad}, we show under which conditions hypergraphs are turn into simplicial complexes or even vanish \textit{tout court}.}
    
    \item The dynamics of the oscillators only depend on the phase differences. This means that there is a rotational symmetry in the reduced model, despite the original system not necessarily having it\footnote{This symmetry is due to the averaging procedure that has removed terms of order $\epsilon^2$ that may break this symmetry.}. {This implies that phase models with rotational symmetry capture the dynamics of any ensemble of weakly coupled oscillators (with close frequencies).}
    
    \item Different coupling schemes in the population of oscillators might yield the same coupling in the phase model; notice that multiple combinations of functions $\bm{Z}$ and $\bm{p_{jkl}}$ may result in the same $\Gamma$ in Eq.~\eqref{eq.gammagen}. Thus, understanding the dynamics of one phase model provides universal information about many different settings of coupled oscillators. 
\end{itemize}

By Fourier expanding $\Gamma_{jkl}$, we can write the phase model, Eq.~\eqref{eq.phaseredhighorder}, as a combination of trigonometric functions,
\begin{equation}\label{eq.genphasemodel}
    \dot{\theta}_j=\omega_j+\epsilon \sum_{k,l=1}^{N}A_{jkl} \sum_{\vec{\ell}} \eta_{\vec{\ell}}\sin( \vec{\ell}\cdot\vec{\theta}+\gamma_{\vec{\ell}}),
\end{equation}
where the vector $\vec{\mathcal{\ell}}=(a,b,-a-b)$, with $a$ and $b$ integers, $\vec{\theta}=(\theta_k,\theta_l,\theta_j)$, and the sums over $\vec{\ell}$ is 
taken over all possible vectors of that form with $a+b\le0$. The computation of $\eta_{\vec{\ell}}$ and $\gamma_{\vec{\ell}}$ for a given oscillator and interaction is derived in~\ref{app.phasetrig}.  From Eq.~\eqref{eq.genphasemodel}, it is easy to extrapolate the generalization to any other higher-order interaction: one needs to consider vectors of the previous form with as many entries as units participating in the interaction.

If 
only the vectors\footnote{{These vectors are such that $a$ and $b$ take values $\pm1$ or $0$ that encode higher-order interactions. Note that the vectors $\vec{\ell}=(1,0,-1)$ and $\vec{\ell}=(0,1,-1)$ yield the classical pairwise Kuramoto interaction.}} $\vec{\ell}=(1,1,-2)$ and $(1,-1,0)$ are considered in Eq.~\eqref{eq.genphasemodel}, we obtain the following phase model:
\begin{equation}
    \dot{\theta}_j=\omega+\epsilon \sum_{k,l=1}^{N}A_{jkl}\left[\kappa_1\sin(\theta_k+\theta_l-2\theta_j+\alpha)+\kappa_2\sin(\theta_k-\theta_l+\beta)\right].
\end{equation}
We can think of this model as the simplest phase model for a general $3$-body coupling.

\subsection{Changes in the topology}\label{sec:degrad}

Previously, we have commented that the phase reduced model can have a different interaction topology to the original setup. In this Section, we explain under which conditions these changes take place. We first comment on the transformation from hypergraphs to simplicial complexes, and then on the disappearance of hyperedges. In Sec.~\ref{sec:SL.degradate}, we show specific examples where these phenomena occur.

\subsubsection{Emergence of simplicial complexes}

From Eq.~\eqref{eq.gammagen}, we observe that, if the phase interacting function $\Gamma_{jkl}$ only depends on one variable (i.e., one phase difference), the hyperedge would be degraded into a pairwise interaction, i.e., a link.
This can only happen if the interaction function is two-body. Notice, however, that we can always write the function $\bm{p}_{jkl}$ as
\begin{equation}\label{eq.psplit}
    \bm{p}_{jkl}(\theta_j,\theta_k,\theta_l)=\bm{p}_{jk}^1(\theta_j,\theta_k)+\bm{p}_{jl}^2(\theta_j,\theta_l)+\bm{p}_{jkl}^3(\theta_j,\theta_k,\theta_l),
\end{equation}
where $\bm{p}_{jk}^1$ and $\bm{p}_{jl}^2$ lack the dependence on one of the variables $\theta_l$ or $\theta_k$, respectively. 
Another possible term $\bm{p}_{kl}(\theta_k,\theta_l)$ encodes a three body interaction (it affects oscillator $j$), so we absorb it in $\bm{p}^3_{jkl}$. We remark that because $\bm{p}$ depends on the limit cycle $\bm{\tilde{X}}_c$, the splitting of Eq.~\eqref{eq.psplit} generally occurs even if 
${\bm g}_{jkl}$ cannot be split into 
${\bm g}_1({\bm X}_j, {\bm X}_k) + {\bm g}_2({\bm X}_j, {\bm X}_l) + {\bm g}_3({\bm X}_j, {\bm X}_k, {\bm X}_l)$;
see Sec.~\ref{sec:SL.degradate} for an example.

The functions $\bm{p}^{1,2}$ 
are computed as:
\begin{eqnarray}
    \bm{p}_{jk}^{1}(\theta_k,\theta_j)&=&\frac{1}{2\pi}\int_0^{2\pi}\bm{g}_{jkl}(\tilde{\bm{X}}_c(\theta_j),\tilde{\bm{X}}_c(\theta_k),\tilde{\bm{X}}_c(\theta_l))d\theta_l \label{eq.plink1},\\
    \bm{p}_{jl}^{2}(\theta_l,\theta_j)&=&\frac{1}{2\pi}\int_0^{2\pi}\bm{g}_{jkl}(\tilde{\bm{X}}_c(\theta_j),\tilde{\bm{X}}_c(\theta_k),\tilde{\bm{X}}_c(\theta_l))d\theta_k \label{eq.plink2}.
\end{eqnarray}

The phase interaction function is then
\begin{equation}
    \Gamma_{jkl}(\Delta\theta_{kj},\Delta\theta_{lj})=\Gamma_{jk}^1(\Delta\theta_{kj})+\Gamma_{jl}^2(\Delta\theta_{lj})+\Gamma_{jkl}^3(\Delta\theta_{kj},\Delta\theta_{lj}),
\end{equation}
with each $\Gamma^j$ obtained by replacing $\bm{p}^j$ in Eq.~\eqref{eq.gammagen}.
This mean that, in addition to the hyperedge given by $\Gamma_{jkl}^3$, the phase reduced model contains two additional links, given by the interaction functions $\Gamma_{jm}^{1,2}(\Delta\theta_{mj})$. If we repeat the procedure for all the hyperedges, we see the original hypergraph is turned into a simplicial complex for the phase reduced model, because for every $3$-body interactions, all the sub-interactions (i.e., the links) are present. 
This occurs under the condition that
$\bm{p}_{jm}^{1,2}\not=0$, which is generally expected in the absence of symmetries. In the general case of $m$-body interactions, the corresponding phase interaction function can be split into terms corresponding to all the $n$-hyperedges with $n\leq m-1$.
This result is particularly significant, as it indicates that phase reduced models with many-body interactions generally interact through simplicial complexes. However, there are some exceptions, as we discuss in the Section below.

\subsubsection{Disappearance of hyperedges}

Under some conditions, a hyperedge can disappear after phase reduction. If the integral in Eq.~\eqref{eq.gammagen} is zero, then the interaction vanishes, despite having a non-zero entry in the adjacency tensor. Although this is not 
generally the case, certain symmetries ensure this circumstance.

\color{black}
Consider an oscillator whose velocity field is antisymmetric with respect to a change of sign of the variables\footnote{One can get to the same conclusions if the symmetry is only { approximately} fulfilled near the limit cycle{, i.e., $\bm{F}(\bm{X})=-\bm{F}(-\bm{X})+\epsilon\bm{G}(\bm{X})$ for $\bm{X}$ close to the limit cycle.}}, that is $\bm{F}(\bm{X})=-\bm{F}(-\bm{X})$. Such symmetry is present in some important systems, such as the van der Pol oscillator~\cite{VdP26,VdP27}, and the rotational symmetry is a particular case of this symmetry. {If the interaction is symmetric (under the change $\bm{X}\to-\bm{X}$), it does not produce any interaction on the phase reduced model 
at the first order, that is, the hyperedges disappear. We prove this statement in what follows.}

Under the considered symmetry of the oscillator, the limit cycle and the phase sensitivity function are odd functions of the phase{, so they can be expressed in Fourier modes:}
\begin{eqnarray}
{\tilde{\bm X}}_c(\theta) &=&\sum_{k=-\infty}^\infty \bm{A}_k  e^{i(2k+1)\theta},\\
\bm{Z}(\theta) &=& \sum_{k=-\infty}^\infty \bm{B}_k e^{i(2k+1)\theta}.
\end{eqnarray}
Notice that even harmonics are not present since they do not preserve this symmetry\footnote{In order to preserve the symmetry, the change $\theta\to\theta+\pi$ has to yield a minus sign. Note that even terms do not change 
their signs under this transformation.}.
Also notice that $\bm{A}_k$ and $\bm{B}_k$ are in general complex vectors, but $\bm{A}_{-k}=\bm{A}_k^{*}$ and $\bm{B}_{-k}=\bm{B}_k^{*}$ to ensure that
$\bm{\tilde{X}}_c$ and $\bm{Z}$ are real.

{ We evaluate the symmetric interaction, $\bm{g}_S(-\bm{X})=\bm{g}_S(\bm{X})$, exerted on the limit cycle to obtain $\bm{p}_S$ expressed in Fourier modes of the phases.}
\begin{equation}
    \bm{p}_S(\theta_1,\theta_2,\dots)=\sum_{n_{1}, ..., n_{N} = -\infty}^{\infty} \bm{a}_{n_{1}, ..., n_{N}}\exp \left( i \sum_{l=1}^{N} n_{l} \theta_l \right),
    \end{equation}
where $\sum_{l=1}^{N} n_{l}$ is even to fulfill the symmetry.

The interaction function is, 
from Eq.~\eqref{eq.gammagen}:
\begin{align}\label{eq.intersym}
&\Gamma_s(\Delta \theta_{1j}, ..., \Delta \theta_{Nj}) \cr
&=
\frac{1}{2\pi}\int_0^{2\pi} \bm{Z}(\phi_j + \varphi) \cdot \bm{p}_S(\phi_1 + \varphi, \dots, \phi_N + \varphi) d\varphi\\
&=\frac{1}{2\pi}\int_0^{2\pi} \sum_{k=-\infty}^\infty \bm{B}_k \cdot  \bm{p}_S(\phi_1 + \varphi, \dots, \phi_N + \varphi) e^{i(2k+1)(\phi_j+\varphi)} d\varphi \notag\\
&= \frac{1}{2\pi}  \sum_{k,n_{1}, ..., n_{N} = -\infty}^{\infty} \bm{B}_k \cdot \bm{a}_{n_{1}, ..., n_{N}} e^{ i (2k+1)\phi_j +i\sum_l n_{l} \phi_l} \int_0^{2\pi}e^{ i (2k+1+\sum_l n_{l}) \varphi}d\varphi. \notag 
\end{align}

Note that the dependence on $\Delta \theta_{kj}$ is implicit in the above expression where $\Delta \theta_{lj} = \theta_l - \theta_j$ and $\phi_{l} = \theta_{l} - \omega t$. 
Because the last integral vanishes since $(2k+1)+\sum_l n_{l}$ is always odd (recall that $\sum_l n_{l}$ is even), the interaction function $\Gamma_s$ is zero, giving no contribution to the phase model.

We remark that we can always split any interaction function into symmetric and antisymmetric parts:
\begin{equation}
    \bm{g}(\bm{X}_1,\bm{X}_2,\dots)=\bm{g}_S(\bm{X}_1,\bm{X}_2,\dots)+\bm{g}_A(\bm{X}_1,\bm{X}_2,\dots),
\end{equation}
where
\begin{eqnarray}
    \bm{g}_S(\bm{X}_1,\bm{X}_2,\dots)&=&\frac{\bm{g}(\bm{X}_1,\bm{X}_2,\dots)+\bm{g}(-\bm{X}_1,-\bm{X}_2,\dots)}{2},\nonumber \\
    \bm{g}_A(\bm{X}_1,\bm{X}_2,\dots)&=&\frac{\bm{g}(\bm{X}_1,\bm{X}_2,\dots)-\bm{g}(-\bm{X}_1,-\bm{X}_2,\dots)}{2}. \nonumber
\end{eqnarray}
In the case of antisymmetric velocity field ${\bm F}$, the previous result shows that only $\bm{g}_A$ would produce any non-vanishing effect in the phase reduced model.

{ From this Section, we stress the importance of the result that systems with the aforementioned
symmetries interacting through symmetric (e.g., quadratic) terms behave as uncoupled when the interactions are weak.} This point is particularly noteworthy since generally only odd higher-order interactions are considered in the literature\footnote{We remind that higher-order coupling functions have to be nonlinear and not additive with respect to all variables~\cite{neuhauser2020multibody,muolo2023turing};
otherwise we can express them as combinations of pairwise interactions.}~\cite{muolo2024phase,gambuzza2021stability}. {Our results justify this absence of even couplings when the units are antisymmetric oscillators. The justification for different types of units may be related to this result and we leave it as an open problem.}

\subsection{Implementation of phase reduction}

In this Section, we summarize the phase reduction scheme in a pedagogical way, so to provide an easy implementation.
The difficulty of the phase reduction procedure lies in the computation of the frequency of the oscillator $\omega$, the phase sensitivity function $\bm{Z}(\theta)$, and the functional form $\bm{\tilde{X}}_c(\theta)$ of the limit cycle; the latter are needed to evaluate $\bm{g}(\bm{X}_j,\bm{X}_k,\bm{X}_l)$ and, thus, to obtain $\bm{p}_{jkl}(\theta_j,\theta_k,\theta_l)$ and $\Gamma$; c.f. Eq.~\eqref{eq.gammagen}.

Those expressions can be obtained analytically if the oscillator has a rotational symmetry, such as the Stuart-Landau oscillator~\cite{Win80,nakao16}, if some perturbative approach is applicable, as for example the van der Pol oscillator~\cite{Leon23a}, or in one-dimensional oscillators such as the leaky-integrate-fire models \cite{Izh07}. In any other case, a phase reduced model must be obtained numerically. The non-existence of exact analytical results does not render the approach useless, since it provides a low dimensional numerical model with clear physical meaning. We describe the numerical implementation in what follows.

Computing numerically the frequency and the limit cycle is straightforward. Given that the limit cycle is stable, the solution would converge to it after some transient. Then, the frequency is computed from the period of the oscillation. In order to obtain the phase sensitivity function, there are mainly two methods. The first one is the direct method, which consists in perturbing the system at each point 
on the limit cycle and measuring how it advances or delays the phase. 
If the mathematical model of the system is known, it is usually more efficient to use the adjoint method~\cite{ermentrout1996type}, where one solves the adjoint equation: 
\begin{equation}
    \omega\frac{d \bm{Z}(\theta)}{d \theta}=-\bm{J}(\theta)^\top \bm{Z}(\theta),
\end{equation}
with ${\bm J}(\theta)$ being the Jacobian of the system evaluated on the limit cycle and $\top$ representing matrix transposition. The phase sensitivity function is the $2\pi$-periodic solution to this equation with the normalization condition 
$\bm{Z}(\theta) \cdot \frac{d\bm{\tilde{X}}_c(\theta)}{d\theta} = 1$~\cite{nakao16,ermentrout1996type,hoppensteadt2012weakly,brown2004phase}. The phase interaction function is computed from Eq.~\eqref{eq.gammagen} and the phase model Eq.~\eqref{eq.phaseredhighorder} is obtained. We remind that because $\Gamma$ is a bi-periodic function of the phase differences, an approximate analytical phase model is obtained by truncating higher-harmonics.

\color{black}

\section{Results on Stuart-Landau oscillators with different higher-order coupling schemes}\label{sec:SL}

In this section, we exemplify the above theory for a population of Stuart-Landau (SL) oscillators. { These examples illustrate the applicability of the phase reduction, and show the advantages of dealing with phase reduced models.}

\subsection{The Stuart-Landau oscillator}

The SL oscillator is a dynamical system that represents the normal form of a supercritical Hopf-Andronov bifurcation:
\begin{equation}
    \dot{W}=(1+i\alpha) W-(1+i\beta)|W|^2W,
\end{equation}
where $W$ is a complex variable, and $\alpha$ and $\beta$ are real parameters. Its natural frequency is $\omega=\alpha-\beta$ and $\beta$ is the non-isochronicity parameter that measures how much the oscillator accelerates when it is displaced from the limit cycle. We can also write this equation in real and imaginary parts ($W=x+iy$):
\begin{equation}\label{eq:SL}
\begin{pmatrix}
\dot{x}\\
\dot{y}
\end{pmatrix}=\bm{F}(x,y)=\begin{pmatrix}
 x- \alpha y-(x^2+y^2)(x-\beta y)\\
 y+ \alpha x-(x^2+y^2)(y+\beta x)
\end{pmatrix}.
\end{equation}
 The SL oscillator has the following exponentially stable limit cycle solution:
 \begin{equation}\label{eq.lcSL}
 {\tilde{\bm{X}}_c(\theta)=}
    \begin{pmatrix}
 \tilde{x}_c(\theta)\\
 \tilde{y}_c(\theta)
\end{pmatrix}=\begin{pmatrix}
 \cos \theta\\
 \sin\theta
\end{pmatrix},
 \end{equation}
where the phase on the limit cycle is $\theta = \omega t$ ($\mbox{mod}\ 2\pi$).

Due to the rotational symmetry of the SL oscillator, the phase sensitivity function $\bm{Z}(\theta)$ can be derived analytically~\cite{Win80,nakao16}. 
\begin{equation}\label{eq:ZSL}
   \bm{Z}(\theta)=
\begin{pmatrix}
 -\sin\theta-\beta\cos\theta\\
 \cos\theta-\beta\sin\theta
\end{pmatrix}.
\end{equation}

In what follows, we consider populations of coupled SL oscillators whose dynamics obeys
    \begin{equation}
        \begin{pmatrix}
\dot{x}_j\\
\dot{y}_j
\end{pmatrix}=\bm{F}(x_j,y_j)+\epsilon \bm{g}_j(x_1,y_1,x_2,y_2,\dots).
    \end{equation}

\subsection{Example I: Stuart-Landau oscillators with all-to-all higher-order interactions}

As a first example, we consider a population of $N$ SL oscillators coupled through all-to-all identical pairwise and $3$-body interactions. The specific interaction term is chosen as:
\begin{equation}\label{eq:alltoall}
    \bm{g}_{j}=\frac{K_1}{N}\sum_{k=1}^N\begin{pmatrix}
 x_k-x_j\\
 0
\end{pmatrix}+
\frac{K_2}{N^2}\sum_{k,l=1}^N\begin{pmatrix}
 x_kx_lx_j-x_j^3\\
 0
\end{pmatrix},
\end{equation}
where the $N$ dependence ensures a well-defined continuum limit $N\to\infty$. Note that only the $x$ components are coupled.  Although these terms do not preserve the rotational symmetry of the model and, hence, cannot be derived by a center manifold reduction, similar couplings have been considered in several studies, both theoretical~\cite{KVK13,muolo2024phase,muolo2024pinning} and experimental \cite{gambuzza2020experimental}.

\begin{figure}[h!]
\centering
\includegraphics[scale=0.6]{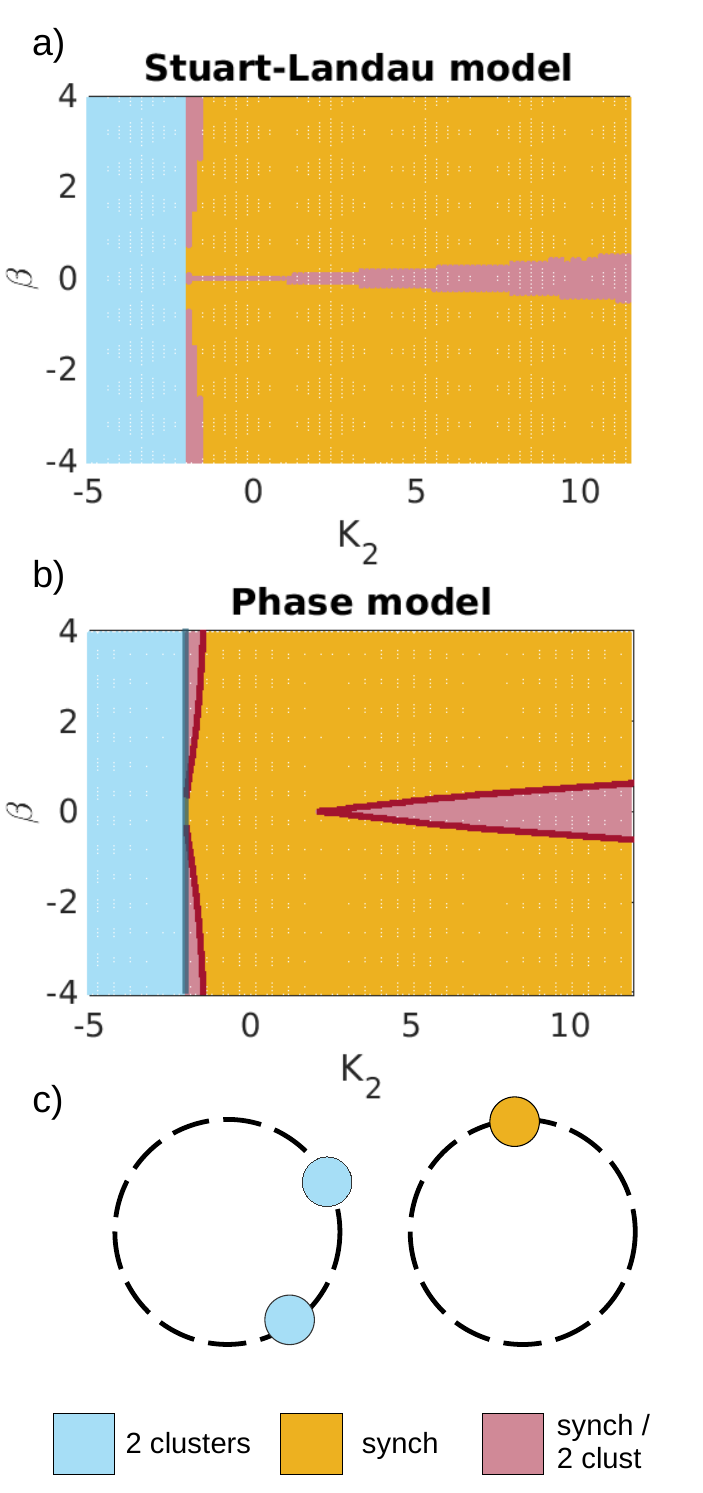}
	\caption
	{Numerically obtained phase diagram for the ensemble of Stuart-Landau oscillators all-to-all coupled through Eq.~\eqref{eq:alltoall} with $\epsilon=0.1$ and $K_1=1$ (a) and for the reduced phase model Eq.~\eqref{eq.phalltoall} with $K_1=1$ (b). In the blue and yellow regions, only two-cluster and fully synchronized states are stable, respectively, while in the red region, we find multi-stability between those states. In panel~(b)
	for the phase model,
we depict 
the theoretical prediction for the stability of synchronization and two-cluster in blue and red, respectively. Panel (c) shows the snapshots of the dynamics for two-cluster and synchronization, respectively.}
	\label{fig:Fig2}
\end{figure}

The absence of rotational symmetry in the coupling makes an analytical study of the original model extremely difficult\footnote{Due to the nonlinear coupling, {analyses such as those carried out in~\cite{nakagawa1993collective,chabanol97} are} not possible. }, 
hence, we perform direct numerical simulations, whose results are shown in Fig.~\ref{fig:Fig2}. 
In Fig.~\ref{fig:Fig2}~(a), we numerically compute the phase diagram of the population of SL oscillators for $\epsilon=0.1$ and $K_1=1$. In the yellow region, fully synchronized state, a solution in which all oscillators form one coherent point-cluster, is the only stable state. Meanwhile, in the blue region, only two-cluster states, solutions in which all oscillators are located in two separate clusters, are stable. In the red regions, we observe multi-stability between synchronization and two-cluster states. 

By performing phase reduction, we can derive a simpler model for which we can obtain analytical results. We compute the phase reduced model of the  ensemble of SL oscillators with the interaction of Eq.~(\ref{eq:alltoall}) {by replacing the coupling function evaluated on the limit cycle, Eq.~\eqref{eq.lcSL}, and phase sensitivity function, Eq.~\eqref{eq:ZSL}, into Eq.~\eqref{eq.gammagen}.}
\begin{equation}
    \dot{\theta}_j=\omega+\frac{\epsilon K_1}{2N}\sum_{k=1}^N [\sin(\theta_k -\theta_j)-\beta \cos(\theta_k -\theta_j)]
    +\frac{\epsilon K_2}{8 N^2}\sum_{k,l=1}^N [\sin(\theta_k +\theta_l -2\theta_j)-\beta\cos(\theta_k +\theta_l -2\theta_j)]
    -\frac{\epsilon \beta K_2}{8 N^2}\sum_{k,l=1}^N [2\cos(\theta_k -\theta_l)-3].
\end{equation}

We can rewrite the model in a more compact form by defining the Kuramoto order parameter $R e^{i\Psi}=\frac{1}{N}\sum e^{i\theta}$, reabsorbing the constant terms by going to a rotating frame of reference $\theta\to\theta-\omega t-3\beta K_2/8$ and rescaling time $t \to \epsilon t$:
\begin{equation}
    \label{eq.phalltoall}
    \dot{\theta}_j=\frac{K_1 R}{2} \left[ \sin(\Psi -\theta_j)-\beta \cos(\Psi -\theta_j)\right]
    +\frac{K_2 R^2}{8} \left[\sin(2\Psi -2\theta_j)-\beta\cos(2\Psi -2\theta_j)
    -2\beta \right].
\end{equation}

A similar model was analyzed in~\cite{leon24higher}, derived from a population of SL oscillators subject to interactions with rotational symmetry. We can follow the analysis of~\cite{leon24higher} to show that, in the continuum limit with $N \to \infty$, the incoherent state, where all oscillators behave independently and $R=0$, is unstable for any value of the parameters $K_1$, $K_2$, and $\beta$. In addition, we can easily compute the stability boundary of the fully synchronized state:
\begin{equation}
    K_2=-2.
\end{equation}
We can also use the analysis of~\cite{KK01,leon24higher} to determine the stability of the 2-cluster states. The analytical stability boundaries of fully synchronized and 2-cluster states are depicted in blue and red, respectively, in the phase diagram of Fig.~\ref{fig:Fig2} (b). In addition, we show the result of numerical simulations with the same color code as in panel (a). We observe perfect agreement between the theory and numerical results. As expected, the numerical phase diagram for the SL oscillators matches the analytical diagram for the phase model (save $\epsilon^2$ corrections). This implies that we can use the phase model Eq.~\eqref{eq.phalltoall} to study the dynamics of the SL oscillators subject to the interaction Eq.~\eqref{eq:alltoall}. Let us remark that such a simple analysis is made possible by the phase reduction and would be very challenging, if not impossible, with other methods, such as the MSF briefly explained in \ref{app:a}.

\subsection{Example II: Stuart-Landau oscillators on a ring-like hypergraph}\label{sec::hyperring}

\begin{figure}[h!]
\centering
\includegraphics[scale=0.5]{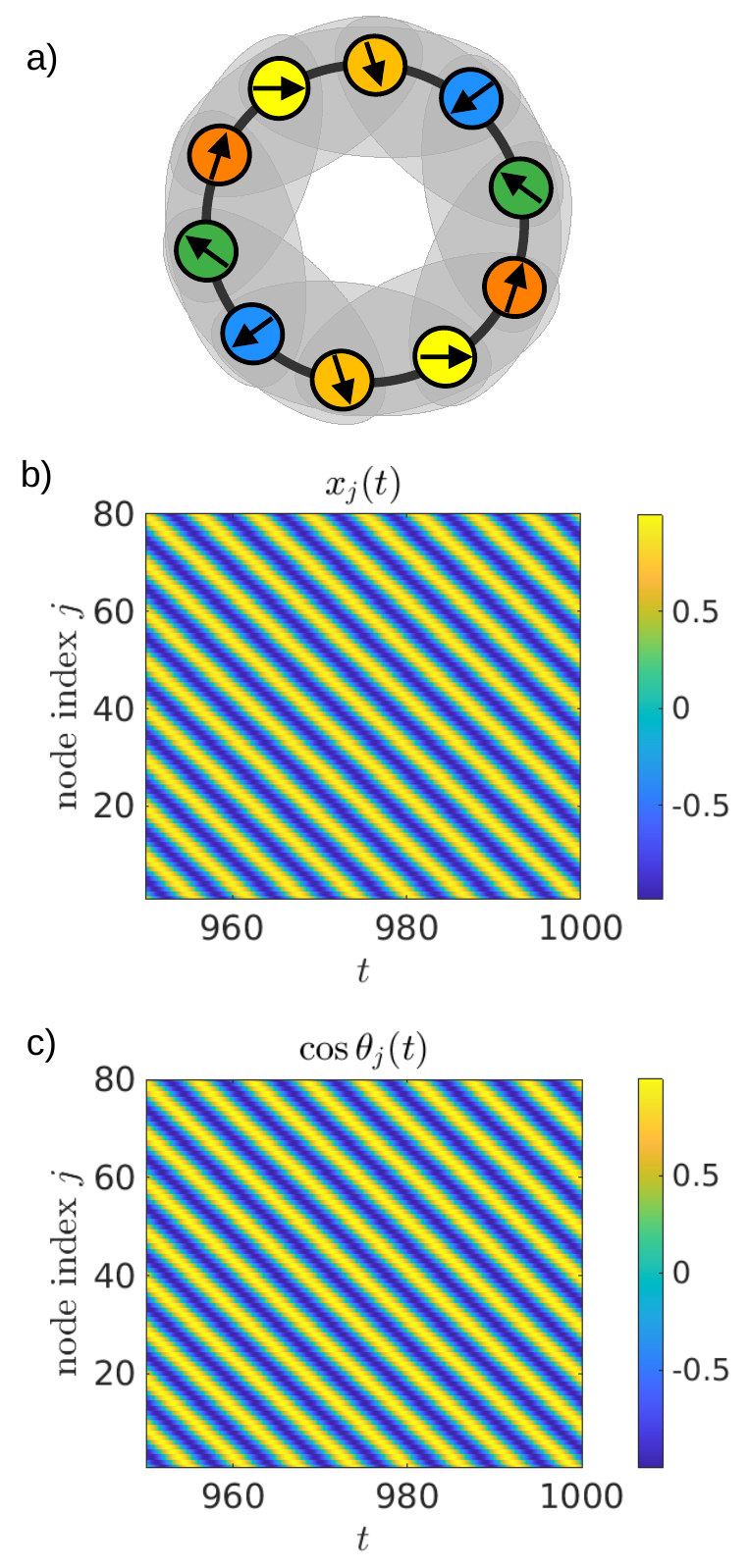}	\caption
	{(a) Schematic representation of the twisted state with two rotations on a ring-like hypergraph: node are circles, links are black lines and hyperedges are gray bubbles. The arrows and colors in the link represent a twisted state in which the phase rotates twice around the ring. (b) Time series of the $x$ component of each SL oscillator obtained by numerically simulating the original model. (c) Time series of $\cos{\theta}$ obtained by numerically simulating the reduced phase model. In (b) and (c), the parameters are $\epsilon=0.1$, $K_1=0.5$, and $K_2=0.5$. The color indicated the value of $x_j$ and $\cos\theta_j(t)$ according to the color bar. }
	\label{fig:Fig3}
\end{figure}

In this second example, we show how phase reduction can be applied not only to the case of all-to-all coupling, but also when the interactions are modeled by a general hypergraph topology. We consider an ensemble of SL oscillators with $\beta=0$, for simplicity, and coupled in a ring-like structure with three-body interactions of the following form:
\begin{equation}
    \bm{g}_{j}=K_1\begin{pmatrix}
 x_{j+1}+x_{j-1}-2x_j\\
 0
\end{pmatrix}+K_2
\begin{pmatrix}
 x_{j+1}x_{j-1}x_j-x_j^3\\
 0
\end{pmatrix}.
\end{equation}
The hypergraph topology is schematically represented in Fig.~\ref{fig:Fig3}~(a). A complete analysis of the dynamics of SL oscillators on such a hypergraph is not possible analytically. However, phase reduction yields a simpler model providing some insights 
into the dynamics. The corresponding phase model {is derived as in the previous section}.
\begin{equation}\label{eq:hypergraph}
    \dot{\theta}_j=\omega+\frac{\epsilon K_1}{2}\left[\sin(\theta_{j-1}-\theta_j)+\sin(\theta_{j+1} -\theta_j)\right]
    +\frac{\epsilon K_2}{8}\sin(\theta_{j+1} +\theta_{j-1} -2\theta_j).
\end{equation}

{ The phase reduced model conserves the original ring-like topology, since the interaction is antisymmetric and the functions $\bm{p}_{mj}^{1,2}=0$, c.f. Eqs.~\eqref{eq.plink1} and \eqref{eq.plink2}.}  Let us observe that such phase model is a particular case of the family studied in~\cite{zhang2024deeper}, after appropriate parameter rescaling. The system exhibits multiple twisted states, in which the phase makes complete rotations along the ring, as schematically represented in Fig.~\ref{fig:Fig3}~(a). The stability of the twisted states increases with the coupling strength $K_2$, as derived in 
\ref{app:c}, although their basins of attraction shrink~\cite{zhang2024deeper}. This knowledge about the phase reduced model can be directly translated to SL oscillators for small $\epsilon$, so similar results are expected. In Fig.~\ref{fig:Fig3}~(b) and (c), we depict the time series of the SL model and the phase reduced model, respectively, for $N=80$ oscillators on the same hypergraph topology and with similar initial conditions. We have fixed $\epsilon=0.1$, $K_1=0.5$, and $K_2=0.5$. Both, the SL model and the phase model, evolve to the same twisted state with eight twists, confirming that the phase reduced model \eqref{eq:hypergraph} is an excellent proxy for the SL model also in this configuration.

\subsection{Example III: phase reduction yielding changes in the topology}\label{sec:SL.degradate}
\begin{figure}[h!]
\centering
\includegraphics[scale=0.5]{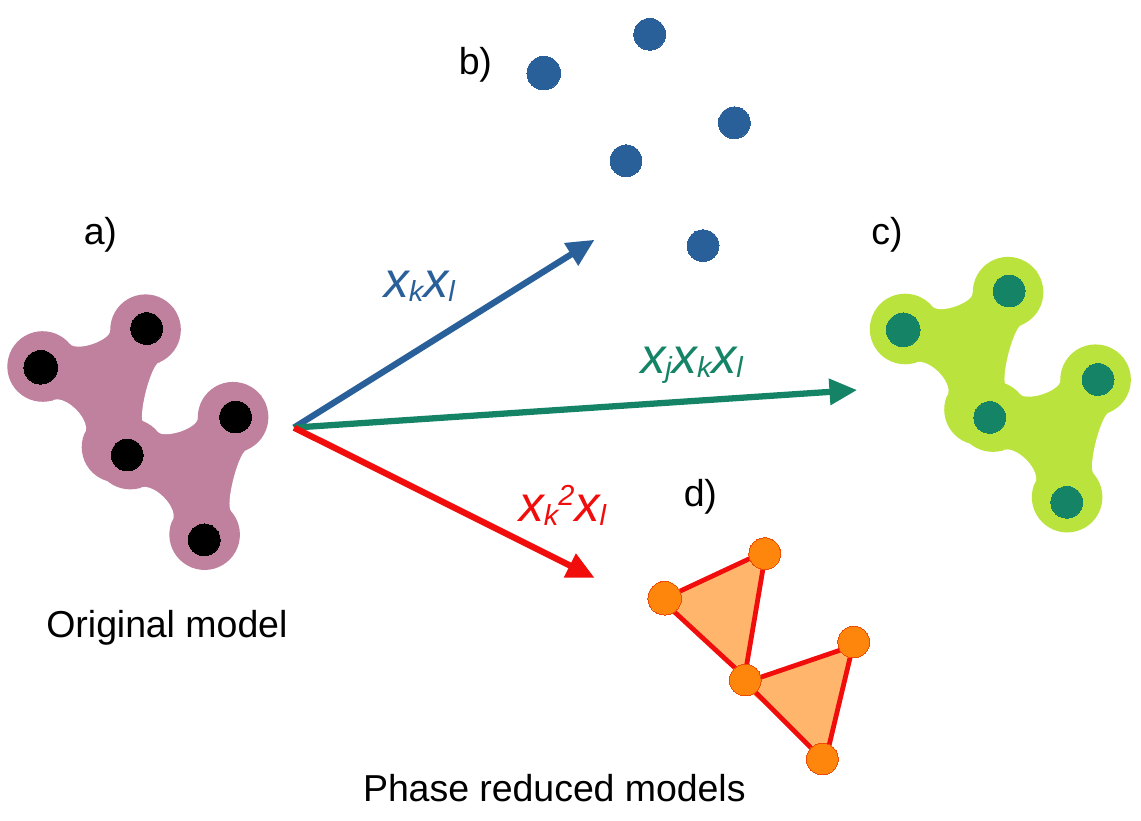}	\caption
	{Effect of phase reduction on the topology (a) for three different interactions, (b) 
	 completely antisymmetric, (c) completely symmetric, (d) symmetric in one of the variables. }
	\label{fig:Fig4}
\end{figure}

In this Section, we show a simple example in which the interaction topology is modified because the three-body coupling either gives rise to additional links or disappears. We consider a population of Stuart-Landau oscillators on a hypergraph with pure $3$-body interactions, encoded by the adjacency tensor $A_{jkl}$. This topology is schematically represented in Fig.~\ref{fig:Fig4} (a).

Three different interaction functions are considered, namely,
\begin{eqnarray}
    {\bm g}_{j}&=&\sum_{kl}A_{jkl} \begin{pmatrix}
 x_{k}x_l-x_j^2\\
 0
\end{pmatrix},\\
    {\bm g}_{j}&=&\sum_{kl}A_{jkl} \begin{pmatrix}
 x_{k}x_lx_j-x_j^3\\
 0
\end{pmatrix},\\
    {\bm g}_{j}&=&\sum_{kl}A_{jkl} \begin{pmatrix}
 x_{k}^2x_l-x_j^3\\
 0
\end{pmatrix},
\end{eqnarray}
which give rise to the following phase models, respectively,
\begin{eqnarray}
    \dot{\theta}_j&=&\omega,\\
    \dot{\theta}_j&=&\omega+\frac{\epsilon}{8}\sum_{kl}A_{jkl}\sin(\theta_k+\theta_l-2\theta_j),\\
    \dot{\theta}_j&=&\omega+\frac{\epsilon}{8}\sum_{kl}A_{jkl}\sin(2\theta_k-\theta_l-\theta_j)+\frac{\epsilon}{4}\sum_{k}B_{jk}\sin(\theta_k-\theta_j),
\end{eqnarray}
where $B_{jk}=\sum_{m}A_{jmk}$, or, equivalently, $B_{jk}=1$ if oscillators $j$ and $k$ originally share a hyperedge and 
$B_{jk}=0$
otherwise. Note that $B_{jk}$ is an adjacency matrix determined from the adjacency tensor $A_{jkl}$, so two oscillators are connected via a link only if they are part of the same $3$-body interaction.

We can see that, from the same initial pure $3$-body topology, three different effective topologies are obtained depending on the interaction scheme. In the first case, because the interaction is symmetric and the SL oscillator has rotational symmetry, the interaction vanishes when phase reduction is performed, as shown in Fig.~\ref{fig:Fig4} (b). In the second case, the topology is preserved, Fig.~\ref{fig:Fig4} (c), as was the case 
in Sec.~\ref{sec::hyperring}. On the other hand, if the oscillators are subject to the last interaction as in \cite{tchinda2025chimera}, we observe that, in addition to the three-body interactions, oscillators sharing a hyperedge are also connected via a link; stated differently, the hypergraph becomes a simplicial complex, Fig.~\ref{fig:Fig4} (d). This is the general expected case when phase reduction is performed, as seen in Sec.~\ref{sec:degrad}. We remark that the pairwise interactions are a result of the phase reduction, since the original interaction 
${\bm g}$ cannot be expressed as a sum of pairwise interactions.

\color{black}
\section{Conclusions and discussion}

In this paper, we have presented a general theory of phase reduction applied to oscillators subject to higher-order interactions. The theory is described for 
any oscillator and three-body interaction {on a hypergraph}, yielding the phase model Eq.~\eqref{eq.phaseredhighorder}. The extension to other higher-order interactions or a mixture of them is straightforward.

{ Phase reduction provides naturally the kinds of higher-order interaction that should be considered in phase models. We observe that, in general, interactions described by simplicial complexes are expected, as each many-body interaction of order $m$ generates all the interactions of lower order $n\leq m$. We also prove that, under certain symmetries of the oscillator,
even interactions do not have any effect on the phase model at the first order. Furthermore, the simplicity of the reduced phase models allows an in-depth analysis, providing information about the non-reduced model, which would be very difficult to obtain by other means. } 

We have exemplified the theory and its results with a paradigmatic case study, the Stuart-Landau oscillator, commonly used because it is a simple and analytically solvable model. The rotational symmetry of the oscillator allows us to obtain analytical phase models. Furthermore, we study { multiple} coupling schemes, showing the usefulness of phase reduction at capturing the dynamics of the original model with a very good approximation. This implies that the conclusions obtained on the more tractable phase models are extensible to the original ensemble of oscillators. Lastly, as phase reduction has recently been extended to take into account stronger couplings \cite{leon19,kurebayashi2013phase,Wilson_2016,shirasaka2017phase}, similar extensions can be considered if higher-order interactions are present. 

Phase reduction has been a powerful tool in the study of oscillatory dynamics in the presence of pairwise interactions, which has justified the use of phase models in many applications, from power grids to neuroscience. We believe that the extension to the case of higher-order interactions can further guide us in understanding how complex interactions shape collective dynamics and provide us with more accurate descriptions of complex oscillatory systems.

\section*{Acknowledgements} 
 I.L. acknowledges support by Grant No.~PID2021-125543NB-I00, funded by MICIU/AEI/10.13039/501100011033 and by ERDF/EU. The work of R.M. is supported by a JSPS postdoctoral fellowship, grant 24KF0211. S.H. acknowledges JSPS KAKENHI 25K15264 for financial support. H.N. acknowledges JSPS KAKENHI 25H01468, 25K03081, and 22H00516 for financial support. The authors are grateful to Timoteo Carletti and Stefanella Boatto for the organization of a symposium on higher-order interactions at the SIAM DS23, from which this work took inspiration. The authors are also grateful to Chris Bick for discussions and feedback. The authors acknowledge Rommel Tchinda Djeudjo for noticing that the nonlinear interaction of Section \ref{sec:SL.degradate} gives rise to both pairwise and higher-order interactions in the reduced model.

\section*{DATA AVAILABILITY}
The data that support the findings of this study are available within the article.

\appendix

\section{All interaction terms} \label{app.phasetrig}

{The general phase model~\eqref{eq.phaseredhighorder} depends on the $\Gamma$ function that is written in terms of a convolution. By use of Fourier modes, we can rewrite such models in terms of trigonometric functions. To do so, we express the limit cycle, phase sensitivity function, and interaction function in terms of Fourier modes of the phase,
\begin{eqnarray}
{\tilde{\bm X}}_c(\theta) &=&\sum_{m=-\infty}^\infty \bm{A}_m  e^{im\theta},\\
\bm{Z}(\theta) &=& \sum_{m=-\infty}^\infty \bm{B}_m e^{im\theta},
\end{eqnarray}
where $\bm{A}_m$ and $\bm{B}_m$ are in general complex vectors, but 
$\bm{A}_{-m}=\bm{A}_m^{*}$ and $\bm{B}_{-m}=\bm{B}_m^{*}$ to ensure that
$\bm{\tilde{X}}_c$ and $\bm{Z}$ are real. If we express the interaction function in terms of the phases, we obtain 

\begin{equation}
\bm{g}_{jkl}(\tilde{\bm{X}}(\theta_j),\tilde{\bm{X}}(\theta_k),\tilde{\bm{X}}(\theta_l))=\bm{p}_{jkl}(\theta_j,\theta_k,\theta_l)=\sum_{n_{j}, n_k, n_l = -\infty}^{\infty} \bm{a}_{n_{j} n_k n_l} e^{i (n_j\theta_j+n_k\theta_k+n_l\theta_l )},
\end{equation}
where $a_{n_j, n_k, n_l}$ is the Fourier coefficient, which yields
\begin{align}
&\Gamma_{jkl}(\Delta \theta_{kj},\Delta \theta_{kl}) \cr
&=
\frac{1}{2\pi}  \sum_{m,n_{j}, n_k, n_l = -\infty}^{\infty}  \bm{B}_m \cdot \bm{a}_{n_{j} n_k n_l} e^{ i [(m+n_j)\phi_j+n_k\phi_k+n_l\phi_l]} \int_0^{2\pi}e^{ i( m+n_j+n_k+n_l) \varphi}d\varphi \notag 
\end{align}
where $\Delta \theta_{km} = \phi_k - \phi_m$.
The last integral is always zero unless the resonant condition $m+n_j= -n_k-n_l$ is fulfilled. Rewriting it in terms of trigonometric functions, we get
\begin{equation}
    \Gamma_{jkl}=\sum_{\substack{n_k, n_l = -\infty \\ n_k+n_l\ge0}}^{\infty} \eta_{n_k n_l}\sin( n_k\Delta\theta_{kj}+n_l\Delta\theta_{lj}+\gamma_{n_k n_l}),
\end{equation}
with 
\begin{equation*}
\eta_{n_k n_l}e^{i\gamma_{n_k n_l}}=\frac{1}{2}\sum_{j=-\infty,\infty}\bm{B}_{-n_j-n_k-n_l}\cdot\bm{a}_{n_{j}n_k n_l},
\end{equation*}
which becomes Eq.~\eqref{eq.genphasemodel} after properly defining the vector $\vec{\ell}$.}

\section{Comparison with the Master Stability Function approach}\label{app:a}

It is also interesting to compare the results that can be obtained through the phase reduction with the Master Stability Function (MSF), a linear stability analysis developed by Fujisaka and Yamada~\cite{fujisaka1983stability}, successively extended on complex networks by Pecora and Carroll, who also named it MSF~\cite{pecora1998master}. The MSF\footnote{Note that the term Master Stability Function is commonly referred solely to the case of strange attractors, while other formulations are used for different attractors, namely, dispersion relation for the case of fixed points and Floquet analysis for the case of limit cycles. Even though the physical meanings can be different, in practice, the procedures are similar and consist in performing a linear stability analysis about the synchronous (i.e., homogeneous) solution and study what will be caused by the perturbation (relax back to the homogeneous state or exponentially diverge) as a function of the coupling. The MSF approach is the most general and, hence, it also includes the latter ones. Lastly, when dealing with coupled Stuart-Landau oscillators, the MSF analysis is sometimes called Benjamin-Feir stability analysis~\cite{nakao2014complex}.} poses several advantages and it is widely used, firstly because it can be applied to coupled systems with different dynamics, namely, fixed points (in this case it is also called dispersion relation)~\cite{NM2010}, limit cycles~\cite{challenger2015turing} and strange attractors~\cite{fujisaka1983stability,pecora1998master}.
Such an approach is very powerful and ductile, however, it poses some limitations: namely, one has to deal with identical oscillators, which need to be close to the synchronous solution, analysis is valid only when the perturbations are infinitesimal\footnote{In fact, the MSF can fail in predicting the effective observed stability when the coupling network is non-normal due to the shrinking of the attraction basin, as it has been shown for the cases of fixed points~\cite{muolo2019patterns}, limit cycles~\cite{muolo2021synchronization}, and strange attractors~\cite{nazerian2024bridging}. In such cases, finite perturbations can make the system unstable even though the MSF analysis predicts stability.} and, above all, the analysis gives no other indication rather than the stability of the fully synchronization. While some limitations can be overcome\footnote{For example, the limitation of dealing with identical systems has been satisfyingly dealt with~\cite{sun2009master,sorrentino2011analysis,ricci2012onset,zhang2017identical}, even for large heterogeneity in the parameters~\cite{nazerian2023synchronization}, and there are specific techniques to characterize states beyond full synchronization~\cite{pecora2014cluster,bayani2024transition}}, it is very difficult to perform such an analysis for systems that are not close to the homogeneous state, which makes the MSF formalism not always suitable to fully characterize systems with rich behaviors and an analysis of the phase reduced system can provide deeper insights.

Of course, also the phase reduction has its limitations. For instance, being a reduction, the phase reduced model does not behave exactly as the full model~\cite{sakaguchi1990breakdown} and this can be visualized also from the numerical results that we show in Fig.~\ref{fig:Fig2} of this paper; further limitations come from the need to deal with a weak coupling and for the units displaying limit cycle solutions (but not necessarily synchronized). However, all these limitations can be overcome to some extent: in fact, one can perform a more accurate phase reduction by not neglecting higher-order terms in the expansion and recover the behavior of the full model~\cite{leon19,bick2024higher}, and scholars have developed methods to perform the phase reduction in the case of strong coupling~\cite{kurebayashi2013phase} and far from the attractors~\cite{Wilson_2016,shirasaka2017phase}.
Nonetheless, phase reduction can be applied only to a system whose attractor is a periodic solution (or quasi-periodic or even a fixed point in some special cases~\cite{mauroy2012use}), while it cannot be applied to the case of chaotic systems. For the latter, the MSF remains the best approach.

In Sec. \ref{sec:SL}, 
the great advantage of phase reduction with respect to the MSF approach
is evident.
The linear stability analysis of the MSF would have given us only the stability region of synchronized state, while no indication on the stability of the incoherent, cluster state or twisted states. In fact, the coupling being all-to-all, methods developed to study cluster synchronization in complex networks~\cite{pecora2014cluster,bayani2024transition} are not applicable and only an analysis of the reduced phase model as carried out above can provide an accurate description of these state, including twisted states~\cite{zhang2024deeper}, or other multi-stability that can emerge~\cite{leon24higher}. Note that, however, the MSF approach can have some numerical advantages with respect to the analysis of the phase model, mainly due to the easy parallelization of the calculations and because it avoids the computation of trigonometric functions.

\section{Twisted state and linear stability on a ring-like hypergraph} \label{app:c}
The phase reduced model on a ring-like hypergraph, Eq.~(\ref{eq:hypergraph}), has a simple twisted state or plane wave solution
\begin{equation*}
\theta_j(t) = \omega t + q j, \quad (j=0, 1, \cdots, N-1),
\end{equation*}
where $q$ is a wavenumber. From the periodic boundary condition $\theta_0(t) + 2 m \pi = \theta_{N}(t)$ where $m$ is an integer, the wavenumber $q$ should satisfy
\begin{equation*}
q = 2 \pi \frac{m}{N}, \quad (m=0, 1, \cdots, N-1).
\end{equation*}
To analyze the linear stability of this twisted state, we add a small variation $\varphi_j(t)$ to $\theta_j(t)$ as
\begin{equation*}
\theta_j(t) = \omega t + q j + \varphi_j(t)
\end{equation*}
and plug into Eq.~(\ref{eq:hypergraph}) to obtain
\begin{align*}
\dot\varphi_j(t) 
&= \frac{\varepsilon K_1}{2} [ \sin( \varphi_{j-1} - \varphi_{j} - q ) + \sin( \varphi_{j+1} - \varphi_{j} + q ) ] + 
\frac{\varepsilon K_2}{8}
\sin( \varphi_{j+1} + \varphi_{j-1} - 2 \varphi_{j} ) ] \cr
&\approx 
\frac{\varepsilon}{8}
\left( 4 K_1 \cos q + K_2 \right) ( \varphi_{j+1} + \varphi_{j-1} - 2 \varphi_j )
\end{align*}
where we linearized in $\varphi_j$ in the second line. Expanding the small variation in a Fourier series as
$\varphi_j(t) = \sum_{k=0}^{N-1} c_k e^{2 \pi i k j / N}$ and plugging it into the linearized equation, the Fourier coefficient $c_k$ obeys
\begin{equation*}
\dot c_k = 
\frac{\varepsilon}{8}
\left( 4 K_1 \cos q + K_2 \right) \left( 2 \cos 2 \pi \frac{k}{N} - 2 \right) c_k, \quad (k=0, 1, \cdots, N-1)
\end{equation*}
Thus, the twisted state of wavenumber $q$ is linearly stable when
\begin{equation*}
4 K_1 \cos q + K_2 > 0 
\end{equation*}
for $k=1, ..., N-1$ since $\varepsilon > 0$. Note that the uniform mode $k=0$ is neutral, due to the rotational symmetry.

\end{document}